\documentclass[aps, prl, twocolumn, superscriptaddress, showpacs, floatfix]{revtex4-1}

\usepackage{graphicx}
\usepackage{amssymb}

\begin{document}

\title{Anomalous Magnetothermopower in a Metallic Frustrated Antiferromagnet}

% \affiliation command applies to all authors since the last \affiliation command. The \affiliation command should follow the other information
\author{Stevan Arsenijevi\'c}
\email{s.arsenijevic@hzdr.de}
\affiliation{High Field Magnet Laboratory (HFML-EMFL), Radboud University, Toernooiveld 7, 6525ED Nijmegen, Netherlands}
\affiliation{Radboud University, Institute of Molecules and Materials, Heyendaalseweg 135, 6525, AJ Nijmegen, Netherlands}
\affiliation{Dresden High Magnetic Field Laboratory (HLD-EMFL), Helmholtz-Zentrum Dresden-Rossendorf, 01328 Dresden, Germany}

\author{Jong Mok Ok}
\affiliation{Department of Physics, Pohang University of Science and Technology, Pohang 790-784, Korea}

\author{Peter Robinson}
\affiliation{High Field Magnet Laboratory (HFML-EMFL), Radboud University, Toernooiveld 7, 6525ED Nijmegen, Netherlands}
\affiliation{Radboud University, Institute of Molecules and Materials, Heyendaalseweg 135, 6525, AJ Nijmegen, Netherlands}

\author{Saman Ghannadzadeh}
\affiliation{High Field Magnet Laboratory (HFML-EMFL), Radboud University, Toernooiveld 7, 6525ED Nijmegen, Netherlands}
\affiliation{Radboud University, Institute of Molecules and Materials, Heyendaalseweg 135, 6525, AJ Nijmegen, Netherlands}

\author{Mikhail I. Katsnelson}
\affiliation{Radboud University, Institute of Molecules and Materials, Heyendaalseweg 135, 6525, AJ Nijmegen, Netherlands}

\author{Jun Sung Kim}
\affiliation{Department of Physics, Pohang University of Science and Technology, Pohang 790-784, Korea}

\author{Nigel E. Hussey}
\email{n.e.hussey@science.ru.nl}
\affiliation{High Field Magnet Laboratory (HFML-EMFL), Radboud University, Toernooiveld 7, 6525ED Nijmegen, Netherlands}
\affiliation{Radboud University, Institute of Molecules and Materials, Heyendaalseweg 135, 6525, AJ Nijmegen, Netherlands}

\date{\today}

\begin{abstract}
We report the temperature $T$ and magnetic field $H$ dependence of the thermopower $S$ of an itinerant triangular antiferromagnet PdCrO$_2$ in high magnetic fields up to 32 T. In the paramagnetic phase, the zero-field thermopower is positive with a value typical of good metals with a high carrier density. In marked contrast to typical metals, however, $S$ decreases rapidly with increasing magnetic field, approaching zero at the maximum field scale for $T >$ 70 K. We argue here that this profound change in the thermoelectric response derives from the strong interaction of the 4$d$ correlated electrons of the Pd ions with the short-range spin correlations of the Cr$^{3+}$ spins that persist beyond the N\'{e}el ordering temperature due to the combined effects of geometrical frustration and low dimensionality.
\end{abstract}


\maketitle

%\section{Introduction}

The interplay between itinerant electrons and even simple magnetic structures can lead to spectacular effects, the giant magnetoresistance seen in magnetic multilayers being arguably the most prominent example~\cite{Baibich}. In geometrically frustrated magnets, complex spin textures that couple to the conduction electrons create an altogether different landscape, where short-range correlations are expected to play a major role. Moreover, since magnetism in metals can be destabilized much more readily than in insulators, magnetic frustration in metallic systems offers a rich playground to search for the emergence of novel transport phenomena. Notable recent examples include the unconventional anomalous Hall effect (AHE) observed in magnetic pyrochlores~\cite{Taguchi,Machida} and the suppression of thermopower in a longitudinal magnetic field in the layered Curie-Weiss metal Na$_x$CoO$_2$~\cite{Wang}.

Despite their obvious potential for new physics, metallic frustrated magnets have been noticeably less studied than their insulating counterparts, largely due to the fact that such materials are rare. Of particular interest are materials in which the conduction electrons and magnetic moments arise from different subsystems. In this context, the quasi-two-dimensional (quasi-2D) antiferromagnet PdCrO$_2$~\cite{Wang,Takatsu,Sun} is somewhat unique. PdCrO$_2$ has a delafossite crystal structure with layers of Pd ions arranged in a triangular lattice stacked between magnetic edge-sharing CrO$_6$ octahedra. The latter contains Cr$^{3+}$ ions with localized (Mott insulating) 3/2 spins which order in the 120$^\circ$ antiferromagnetic (AFM) structure below $T_N =$ 37.5 K~\cite{Mekata,TakatsuPRB}. The frustration parameter $f$, defined as an absolute ratio of the Weiss temperature $\Theta_W$ and the ordering temperature $T_N$, is around 13 for PdCrO$_2$, indicating a high level of frustration \cite{Mekata,TakatsuPRB,Doumerc}. According to band structure calculations~\cite{Sobota}, angle-resolved photoemission \cite{Sobota} and quantum oscillation (QO) studies \cite{Ok,Hicks}, the Fermi surface (FS) of PdCrO$_2$, in the paramagnetic (PM) phase, is identical to that of the nonmagnetic analog PdCoO$_2$, and thus is derived uniquely from the 4$d$ electrons on the Pd site. PdCrO$_2$ also draws special interest because it too exhibits an unconventional AHE, i.e. one that does {\it not} scale with its magnetization ~\cite{Takatsu,Ok}.

Here, we report the discovery of a new feature in the transport properties of PdCrO$_2$, namely a strong magnetothermopower (MTP) at elevated temperatures. In a transverse field, $S$ exhibits a marked decrease which for $T >$ 70 K, reaches 100 \% of the zero-field value at $\mu_0H =$ 30 T. The suppression is reminiscent of that first reported in Na$_x$CoO$_2$ ($x =$ 0.7)~\cite{Wang} and attributed to a lifting of the spin degeneracy of the large spin entropy of the mobile Co$^{4+}$ spins that gives rise to its enhanced thermopower \cite{Terasaki}. We argue here however, that the suppression of $S(B)$ in PdCrO$_2$ is distinct from that observed in Na$_x$CoO$_2$ and signifies instead a novel magnon drag contribution to the thermopower that persists far beyond $T_N$ due to the highly frustrated short-range spin correlations on the Cr sublattice.

Single crystals of PdCrO$_2$ of typical dimensions 1$\times$0.4$\times$0.2~mm$^3$ were grown by a flux method, as described in Refs.~\cite{TakatsuCG, Ok}. Details of our thermoelectric measurements can be found in the Supplemental Material (SM)~\cite{SuppInf}. In all measurements reported here, the magnetic field is oriented perpendicular to the thermal gradient and to the highly conducting planes.

\begin{figure}
\centering
\includegraphics[width=1.0\linewidth]{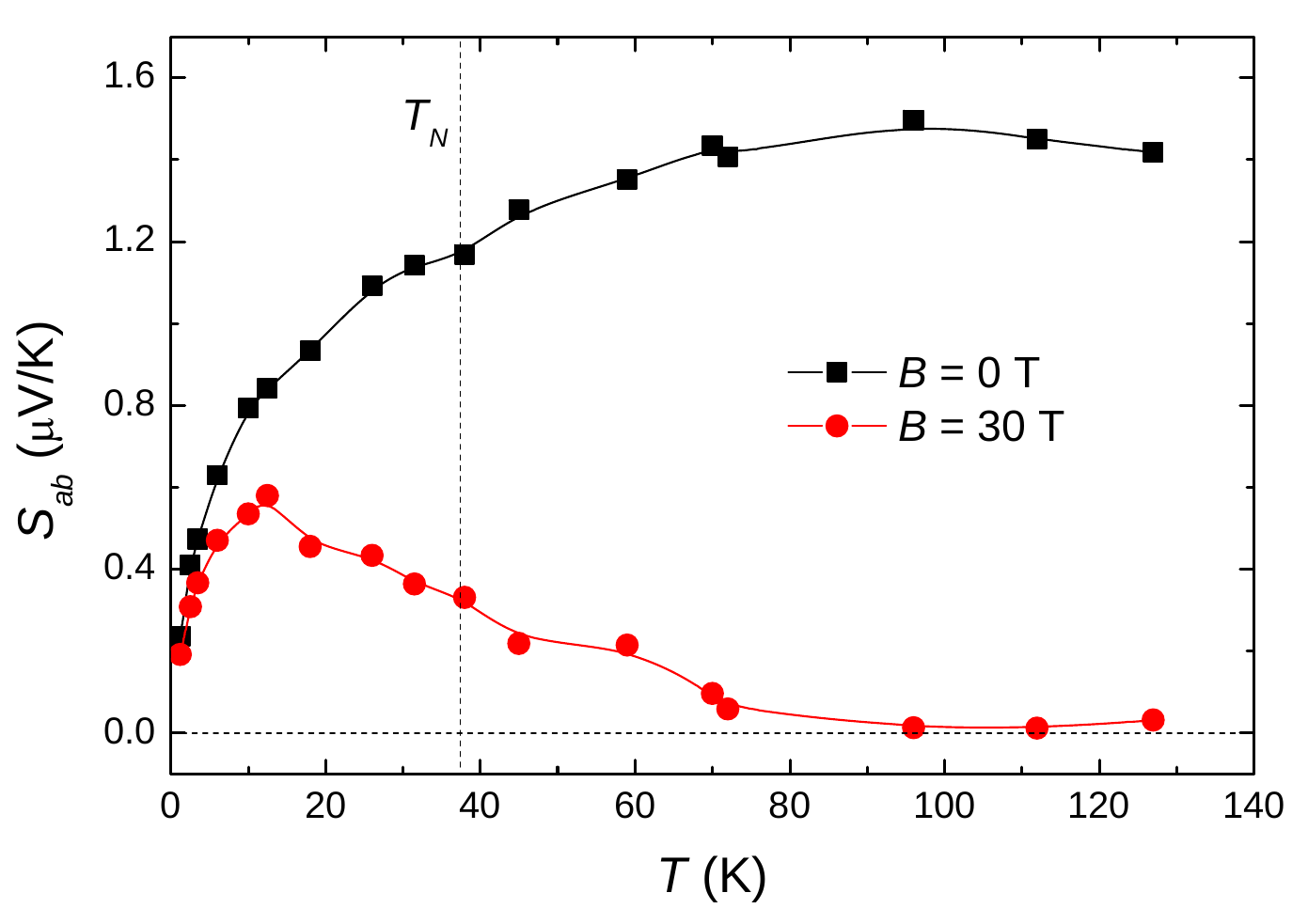}
\caption{(Color online). Temperature dependence of the in-plane thermopower $S_{ab}$ of PdCrO$_2$ at zero and high magnetic field $B$ $\parallel$ $c$ shows an almost complete field suppression for $T >$ 70 K. The magnetic ordering transition at $T = T_N$ is indicated by a vertical dashed line.}
\label{fig:SvsT}
\end{figure}

The key finding of our study is the effect of a magnetic field on the thermoelectric response of PdCrO$_2$. This is summarized in Fig.~\ref{fig:SvsT} where the in-plane thermopower $S_{ab}(T)$ in zero field (solids black squares) is compared with that obtained in an applied field of 30~T (solid red circles) (see Supplemental Material for $S_{ab}(T)$ data at intermediate fields~\cite{SuppInf}). In zero field, $S_{ab}$ has a small positive value less than 2 $\mu$V/K --- typical of good metals --- while the sign and order of magnitude of $S_{ab}$ at $T =$ 130 K are similar to those found in PdCoO$_2$~\cite{Daou}. In a large magnetic field, as summarized in Fig. \ref{fig:SvsT}, the thermopower is almost completely suppressed for $T >$ 70 K. Lowering the temperature towards the magnetically ordered phase reduces the magnitude of the field-suppression of $S_{ab}$.

Such a large suppression of $S$ is not expected in a conventional metal, where the field has a negligible effect on the relative spin-up and spin-down populations of electrons and their entropic current \cite{Ziman}. In strongly correlated electron systems, however, the spin degrees of freedom can give a large contribution to $S$ according to Heikes formula $S = \mu/eT = S_E/e$, where $\mu$ is the chemical potential, and $S_E$ is the entropy per charge carrier~\cite{Chaikin}. Since $S_E$ depends on the spin and configuration degeneracies, the spin entropy term can raise $S$ to order $k_B/e$ and subsequently be suppressed to zero in a magnetic field by a lifting of the spin degeneracy. Such an effect was observed first in Na$_x$CoO$_2$~\cite{Wang} where the field dependence of $S(B)$ for different $T$ was found to be consistent with the variation of the residual spin entropy $S_E(B,T)$ for noninteracting spins in a magnetic field.

\begin{figure}
\centering
\includegraphics[width=1.0\linewidth]{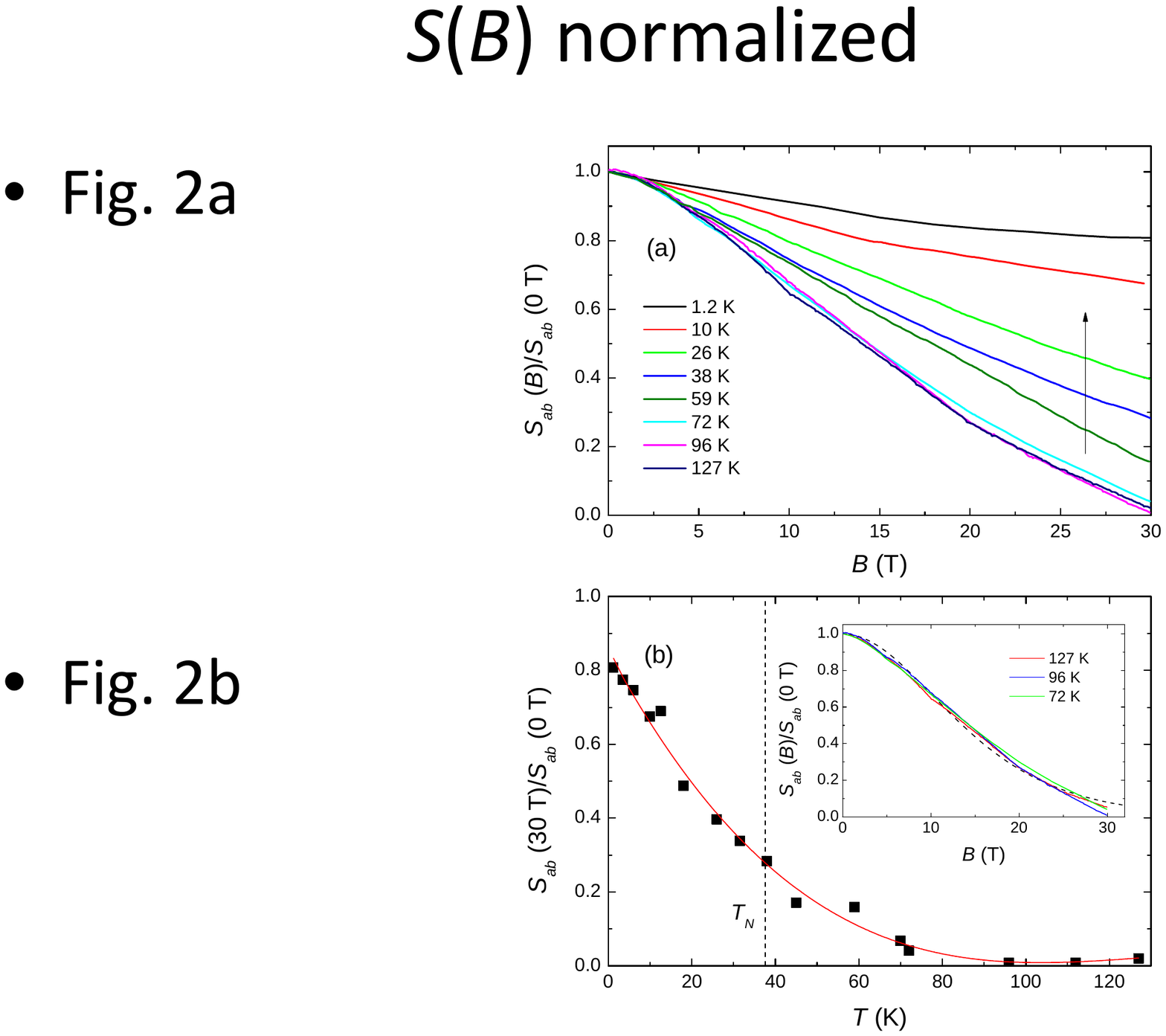}
\caption{(Color online). (a) Magnetic field dependence of $S_{ab}$ as a function of temperature. The suppression of the normalized thermopower by an out-of-plane magnetic field decreases with lowering temperature, as indicated by the arrow. (b) $T$ dependence of the normalized field change of $S_{ab}$ in an applied field of 30~T. Line is a guide to the eye. The inset shows the comparison of a modeled spin entropy in field from Ref.~\cite{Wang} (dashed curve) and the normalized $S_{ab}(B)$ for $T >$ 70~K. While the form of the suppression is consistent with the model, its magnitude clearly does not scale with $\mu_0H/k_BT$.}
\label{fig:SvsB}
\end{figure}

The field dependence of $S_{ab}(B)$ in PdCrO$_2$, normalized to its zero-field value, is shown in Fig.~\ref{fig:SvsB}a for constant temperature field sweeps over a range of temperatures 1.2 K $\leq T \leq$ 130 K. As shown in the inset of Fig.~\ref{fig:SvsB}b, the form of the field suppression is qualitatively similar to that found in Na$_x$CoO$_2$~\cite{Wang}. Indeed, since both transition-metal oxides form a layered triangular lattice of localized, but frustrated magnetic moments which interact with the conducting $d$ electrons, it might be tempting to assign the same origin to the field suppression of $S(B)$ in both cases. Closer inspection, however, reveals some important differences between the compounds and their thermoelectric response that suggest otherwise.

First, in Na$_x$CoO$_2$ ($x$ = 2/3), one third of the Co ions are in a Co$^{4+}$ configuration, giving rise to a band of mobile but AFM-coupled charges (with spin $s = 1/2$) moving through a magnetically inert background of $s =$~0 moments localized on the Co$^{3+}$ sites. Wang {\it et al.} showed that the spin entropy term associated with these mobile spin excitations accounts for almost all of $S$ at 2~K and a dominant fraction at 300 K~\cite{Wang}. In PdCrO$_2$, on the other hand, the sea of conduction electrons is comprised uniquely from the 4$d$/5$s$ states of the Pd ions, as shown convincingly in recent QO studies~\cite{Hicks}, while the Cr$^{3+}$ states are (Mott) insulating and therefore cannot, by themselves, respond to a thermal gradient. Thus, while there is significant spin entropy in the system, it does not contribute to the thermopower of PdCrO$_2$, and correspondingly, there is no enhancement in $S_{ab}$.

Second, the temperature evolution of the suppression is markedly different in the two compounds. In Na$_x$CoO$_2$, the relative suppression of $S$ in field grows with decreasing $T$ as the contribution of the spin entropy term becomes ever more dominant. In PdCrO$_2$, by contrast, the field suppression becomes more pronounced with {\it increasing} temperature. Hence, the $H/T$ scaling observed in Na$_x$CoO$_2$~\cite{Wang} fails in PdCrO$_2$ [see inset to Fig.~\ref{fig:SvsB}(b)]. As illustrated in the main panel of Fig.~\ref{fig:SvsB}(b), where the ratio $S(B =$ 30 T)/$S(B =$ 0 T) is plotted, a total suppression of $S_{ab}$ is only observed for $T \geq$ 2$T_N$. Below this temperature scale, $S$(30 T) remains finite and grows in magnitude with decreasing temperature. 

%Finally, in Na$_x$CoO$_2$, the field suppression is only partial in the transverse field orientation, complete in the longitudinal configuration, while in PdCrO$_2$, it is complete in the transverse configuration (at elevated temperatures).

Large magnetothermopower is often observed when the zero-field thermopower itself has an enhanced value, e.g. as found in semiconductors, and is commonly attributed to the effects of a magnetic field on the respective mobilities of the electron and hole carriers \cite{Wang14}. A sizeable MTP is also found in systems exhibiting a large magnetoresistance~\cite{Sun03}. Neither of these scenarios apply to PdCrO$_2$. As mentioned above, the zero-field thermoelectric response in PdCrO$_2$ is that of a single-band metal with a high carrier density. Moreover, at these elevated temperatures where the suppression of the thermopower is most complete, the in-plane magnetoresistance of PdCrO$_2$ is very small, of order 5$\%$ or less in a field of 30 T~\cite{SuppInf}. Thus, it would appear that the large MTP in PdCrO$_2$ stems from a different origin. Before discussing this in more detail, however, we first complete the summary of our experimental findings, some of which have an important bearing on this discussion.

\begin{figure}
\centering
\includegraphics[width=1.0\linewidth]{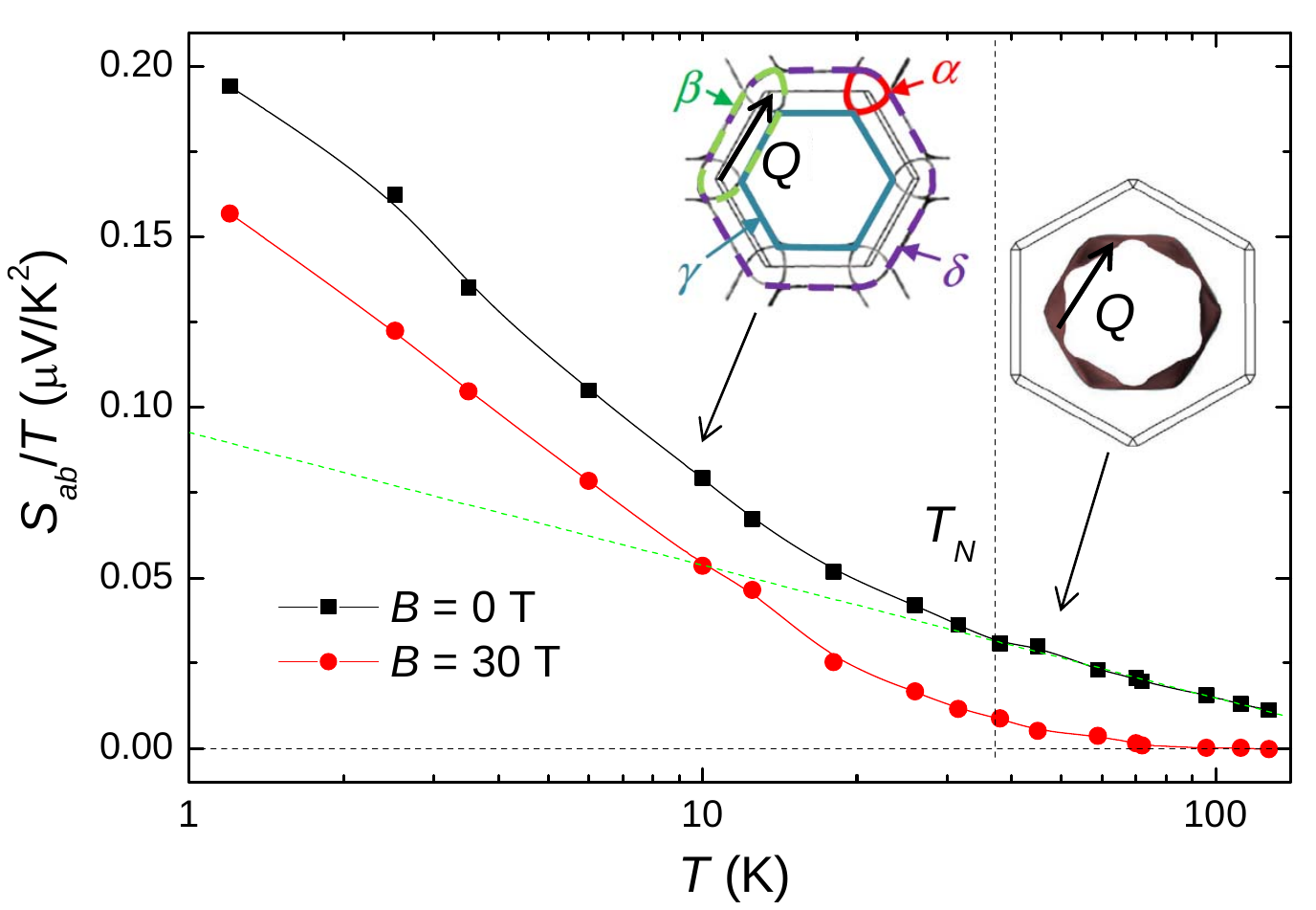}
\caption{(Color online). $S_{ab}/T$ as a function of temperature, plotted on a semilogarithmic scale, in zero field and at $B$ = 30 T. The dashed line is a logarithmic fit to $S_{ab}/T$ above $T_N$ that is associated with the scattering of itinerant electrons on spin fluctuations with a wave vector $Q$~\cite{KimPepin}, as indicated in the Fermi surface model shown in the right inset~\cite{Billington}. Below $T_N$, the enhancement is attributed to a dominant contribution from the $\alpha$ pocket in the reconstructed zone, labeled in the left inset~\cite{Ok}.}
\label{fig:SdTvsT}
\end{figure}

Figure~\ref{fig:SdTvsT} shows the variation of $S_{ab}/T$ in PdCrO$_2$ between 1 K and 130 K, both in 0 T and 30 T, on a semilogarithmic scale. Over the entire temperature range, $S_{ab}/T$ in zero field increases with decreasing temperature. Above $T_N$, $S_{ab}/T$ appears to exhibit a logarithmic enhancement which we attribute to the scattering of electrons on short-range magnetic correlations in the PM phase that are also presumed to be responsible for changes in the zero-field~\cite{TakatsuPRB} or low-field~\cite{DaouPRB} transport properties. The established FS topology of PdCrO$_2$ makes the electron states highly sensitive to scattering processes, either from magnons or spin fluctuations, associated with the AFM wave vector $Q$ (illustrated in the right inset of Fig.~\ref{fig:SdTvsT})~\cite{Billington}. In an isotropic antiferromagnet, the scattering probability $W_q$ with momentum-transfer $q$, while vanishing for $q \rightarrow 0$, is formally divergent for $q \rightarrow Q$, $W_q \propto 1/\omega_q \propto |q-Q|^{-1}$~\cite{Irkhin00}. Above a threshold temperature $T^{*} \sim$ 20--50 K (see Ref.~\cite{SuppInf} and below for a fuller description), ``singular'' scattering off such FS hot spots can lead to a contribution to $S \propto \ln T^{*} / T$~\cite{SuppInf}, qualitatively consistent with what is found here in zero field above $T_N$ (Fig.~\ref{fig:SdTvsT}).

\begin{figure}
\centering
\includegraphics[width=1.0\linewidth]{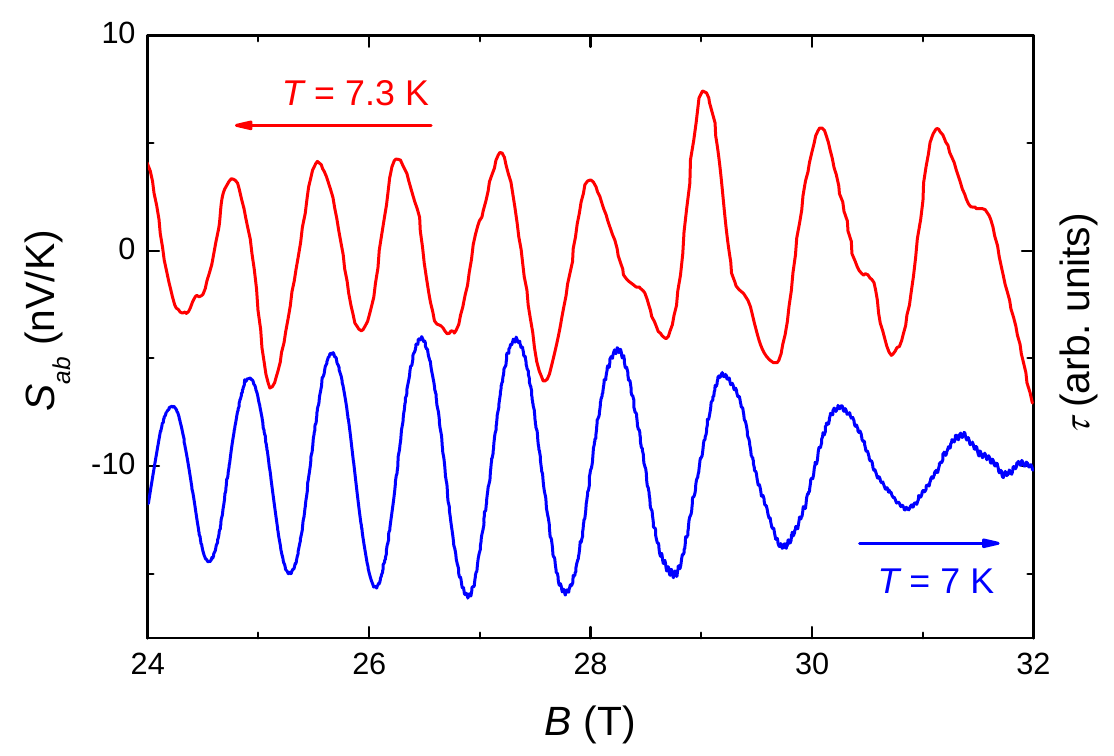}
\caption{(Color online). Slow quantum oscillations in $S_{ab}$ (obtained by subtracting a polynomial background) and in magnetic torque (obtained previously~\cite{Ok}) originating from a small electron pocket in the reconstructed Fermi surface.}
\label{fig:SQO}
\end{figure}

At lower temperatures, where the Cr$^{3+}$ spins order, $S_{ab}/T$ increases more rapidly as the conduction electrons finally undergo FS reconstruction~\cite{Ok, Hicks} (note the upward deviation from the dashed line in Fig.~\ref{fig:SdTvsT} below $T_N$). The existence of small pockets is confirmed by the observation of QO in the high-field thermopower below $T_N$, shown in Fig.~\ref{fig:SQO}. Both the QO frequency $F$ = 710 T and the obtained mass  $m^{*} =$ 0.31(3)~$m_e$~\cite{SuppInf} agree with those previously attributed to the small $\alpha$ pocket in the reconstructed zone~\cite{Ok, Hicks}. Assuming that the $\alpha$ pocket is the dominant contribution to $S_{ab}$ in the low-$T$ limit (given that it has the smallest $T_F$ = 3000 K), we use a simple Drude model to estimate $S_{ab}/T$ = 140 nV/K , comparable to the measured value at 1.2 K of 200 nV/K. Thus, we can attribute the additional enhancement of $S_{ab}/T$ below $T_N$ to FS reconstruction. In contrast to the thermoelectric response of other systems (such as the parent pnictide BaFe$_2$As$_2$~\cite{Stevan}) that undergo a magnetic zone folding, the change in $S/T$ here is very gradual.

%The identical periodicity of the QO found in $S_{ab}/T$ and magnetic torque confirms that in-plane transport detects the same FS structure as seen in the bulk magnetization.

Let us now turn to discuss the origin of the unusual field dependence of $S_{ab}$ in PdCrO$_2$. The observed FS reconstruction indicates strong coupling between the itinerant electrons and the local moments on the Cr$^{3+}$ sublattice. According to specific heat and susceptibility data~\cite{TakatsuPRB}, there is a broad region in temperature above $T_N$, extending up to 150 K (i.e. 4$T_N$), in which short-range correlations among the frustrated spins persist. This is also confirmed by the observation of diffuse magnetic scattering in Refs.~\cite{Mekata, TakatsuPRB, Billington}, discussed in more detail in the Supplemental Material. Such an extended range of critical behavior is a feature of quasi-2D triangular \cite{Hirakawa, Olariu, Alexander} and kagome \cite{Grohol} lattice magnets.
%The in-plane resistivity also exhibits a sharp drop below $T_N$, due to a freezing out of the spin fluctuations, and a $T$-sublinear dependence above which has been ascribed to a gradual reduction in magnetic disorder scattering with increasing temperature \cite{Takatsu}. 
%The existence of short-range spin correlations developing in the paramagnetic phase above $T_N$ is also confirmed by the observation of diffuse magnetic scattering in Ref. \cite{Mekata, Takatsu} and most recently in Ref. \cite{Billington}.

Short-range magnetic order in the PM phase, with a correlation length $\xi \gg a$, the lattice parameter, can in principle persist up to a temperature scale of the AFM exchange energy, i.e., $T \sim J \gg T_N$. As was shown in Ref.~\cite{Irkhin91}, the character of electron-magnon interactions does not change markedly at $T_N$ assuming that $\xi \gg k_F^{-1}$ which for metals is the same as $T < J$. Below a threshold temperature $T^{*} \sim (\Delta / E_F)J \sim$ 20--50 K, where $\Delta$ is the AFM gap and $E_F$ the Fermi energy, the thermopower will be dominated by the usual diffusive term that is weakly dependent on magnetic field. Above $T^{*}$, however, the strong electron-magnon interaction with $q \rightarrow Q$ will become relevant~\cite{Irkhin00}, giving rise to a magnon-drag contribution $S_g$ to the total thermopower, which can become significant provided that the magnon-electron scattering is comparable to the scattering of magnons by defects, which seems a reasonable assumption in a clean, stoichiometric PdCrO$_2$. The magnon drag term will be strongly dependent on the spin-wave spectrum which is also very sensitive to magnetic field~\cite{Starykh}. According to the theory described in Ref.~\cite{Zhitomirsky}, AFM magnons become unstable with respect to two-magnon decay processes at high enough magnetic fields. This effect leads to a strong suppression of magnon drag and its contribution to the thermoelectric power which explains qualitatively the dramatic growth of the MTP at higher temperatures. This picture should be contrasted with the typical magnon-drag scenario proposed in AFM and FM metals where a strong MTP is only seen {\it below} the magnetic transition and is absent above it~\cite{Bhandari, Grannemann, Costache, Caglieris}. The persistence of such a term in PdCrO$_2$ above $T_N$ is then a direct consequence of the highly frustrated nature of the magnetic order arising from the combined effects of geometrical frustration and low dimensionality.

Another interesting possibility for the large field suppression of $S$ at elevated temperatures is a reduction of the scattering amplitudes related to the interaction between the itinerant electrons and the emergent spin chirality. At temperatures above $T_N$, the fluctuating Cr spins are easily aligned in a field, forming new spin textures which give rise, owing to their triangular arrangement, to a finite scalar spin chirality. Indeed, analysis of electron spin resonance experiments in PdCrO$_2$ have shown evidence for spin relaxation processes involving $Z_2$ vortices associated with chiral fluctuations of the 120$^\circ$ spin structure extending up to 300 K~\cite{Hemmida}. Moreover, the unconventional AHE observed in PdCrO$_2$ above $T_N$ has been attributed to a strong coupling of the itinerant electrons to the emergent field-induced spin chirality~\cite{Ok}. The evolution of the field-induced suppression of the thermopower in PdCrO$_2$ suggests that both phenomena may be linked to the same physics, since below $T_N$, where the 120$^\circ$ spin structure becomes increasingly more resilient to an applied field, both the MTP and AHE are correspondingly reduced. The form of the in-plane resistivity in PdCrO$_2$ above $T_N$ has also been attributed to magnetic scattering off short-range magnetic correlations among the frustrated spins~\cite{TakatsuPRB}. It is therefore tempting to attribute the suppression of $S_{ab}(B)$ to a similar effect. For the effect to grow with increasing temperature, however, either the coupling of the conduction electrons to the underlying spin textures would have to become stronger, or the induced chirality more pronounced as $T$ increases.

In summary, we have uncovered a marked suppression of the in-plane thermopower of the metallic frustrated antiferromagnet PdCrO$_2$ in high magnetic fields up to 32 tesla. Certain features of the thermoelectric response suggest that this suppression is in the metallic, rather than in the spin entropic contribution to $S_{ab}$. The temperature evolution of the suppression, in particular, implies a dominant magnon-drag contribution that persists far beyond $T_N$ due to the thermally-robust interaction between the conduction electrons and the short-range magnetic correlations.

\begin{acknowledgments}

The authors thank C. Proust and N. Shannon for stimulating discussions and acknowledge the support of the HFML-RU/FOM and HLD/HZDR, members of the European Magnetic Field Laboratory (EMFL). The work at POSTECH was supported by  the NRF through SRC (Grant No. 2011-0030785), Max Planck POSTECH/KOREA Research Initiative (Grant No. 2011-0031558) Programs, and also by IBS (No. IBSR014-D1-2014-a02). A portion of this work was performed with the help of Dr.~E.~S. Choi at the National High Magnetic Field Laboratory, which is supported by National Science Foundation Cooperative Agreement No. DMR-1157490, the State of Florida, and the U.S. Department of Energy.

\end{acknowledgments}

\end{document}

\title{Supplementary information: Anomalous Magnetothermopower in a Metallic Frustrated Antiferromagnet}


\author{Stevan Arsenijevi\'c}
\affiliation{High Field Magnet Laboratory (HFML-EMFL), Radboud University, Toernooiveld 7, 6525ED Nijmegen, Netherlands}
\affiliation{Radboud University, Institute of Molecules and Materials, Heyendaalseweg 135, 6525, AJ Nijmegen, Netherlands}
\affiliation{Dresden High Magnetic Field Laboratory (HLD-EMFL), Helmholtz-Zentrum Dresden-Rossendorf, 01328 Dresden, Germany}

\author{Jong Mok Ok}
\affiliation{Department of Physics, Pohang University of Science and Technology, Pohang 790-784, Korea}

\author{Peter Robinson}
\affiliation{High Field Magnet Laboratory (HFML-EMFL), Radboud University, Toernooiveld 7, 6525ED Nijmegen, Netherlands}
\affiliation{Radboud University, Institute of Molecules and Materials, Heyendaalseweg 135, 6525, AJ Nijmegen, Netherlands}

\author{Saman Ghannadzadeh}
\affiliation{High Field Magnet Laboratory (HFML-EMFL), Radboud University, Toernooiveld 7, 6525ED Nijmegen, Netherlands}
\affiliation{Radboud University, Institute of Molecules and Materials, Heyendaalseweg 135, 6525, AJ Nijmegen, Netherlands}

\author{Mikhail I. Katsnelson}
\affiliation{Radboud University, Institute of Molecules and Materials, Heyendaalseweg 135, 6525, AJ Nijmegen, Netherlands}

\author{Jun Sung Kim}
\affiliation{Department of Physics, Pohang University of Science and Technology, Pohang 790-784, Korea}

\author{Nigel E. Hussey}
\affiliation{High Field Magnet Laboratory (HFML-EMFL), Radboud University, Toernooiveld 7, 6525ED Nijmegen, Netherlands}
\affiliation{Radboud University, Institute of Molecules and Materials, Heyendaalseweg 135, 6525, AJ Nijmegen, Netherlands}

\maketitle

\renewcommand{\bibnumfmt}[1]{[S#1]}
\renewcommand{\citenumfont}[1]{S#1}
\renewcommand{\thefigure}{S\arabic{figure}}

\textbf{Experimental details.} Single crystals of PdCrO$_2$ of typical dimensions 1 x 0.4 x 0.2 mm$^3$ were grown by a flux method, as described in Ref.~\cite{TakatsuCG, Ok}. The experiments were carried out at the High Field Magnetic Laboratory (HFML) in Nijmegen using a specially built sample-holder designed to be incorporated into one of the He-4 flow cryostats available at HFML. The cryostat themselves are designed for measurements in the temperature range between 2 K and 100 K. The observed thermopower suppression in magnetic field motivated us to extend the operating range of this system to around 130 K, more than three times $T_N$ of PdCr$O_2$. The sample holder has a one-heater-two-thermometers configuration with RuO$_2$ bare chip resistors used for thermometry. The thermoelectric voltage was measured with phosphor-bronze lead wires for the majority of the field sweeps because of their low background contribution. Exceptionally, in the search for Shubnikov-de-Haas oscillations in the thermopower, gold wires were used for their smaller resistance values. The thermoelectric voltage was captured using a low-noise analog nanovoltmeter connected to twisted pairs of signal wires leading to the cold point of the holder in order to inhibit various noise sources. In all measurements reported in the main text, the magnetic field is oriented perpendicular to the thermal gradient and to the highly conducting planes.

\begin{figure}[!thb]
\centering
\includegraphics[width=1.0\linewidth]{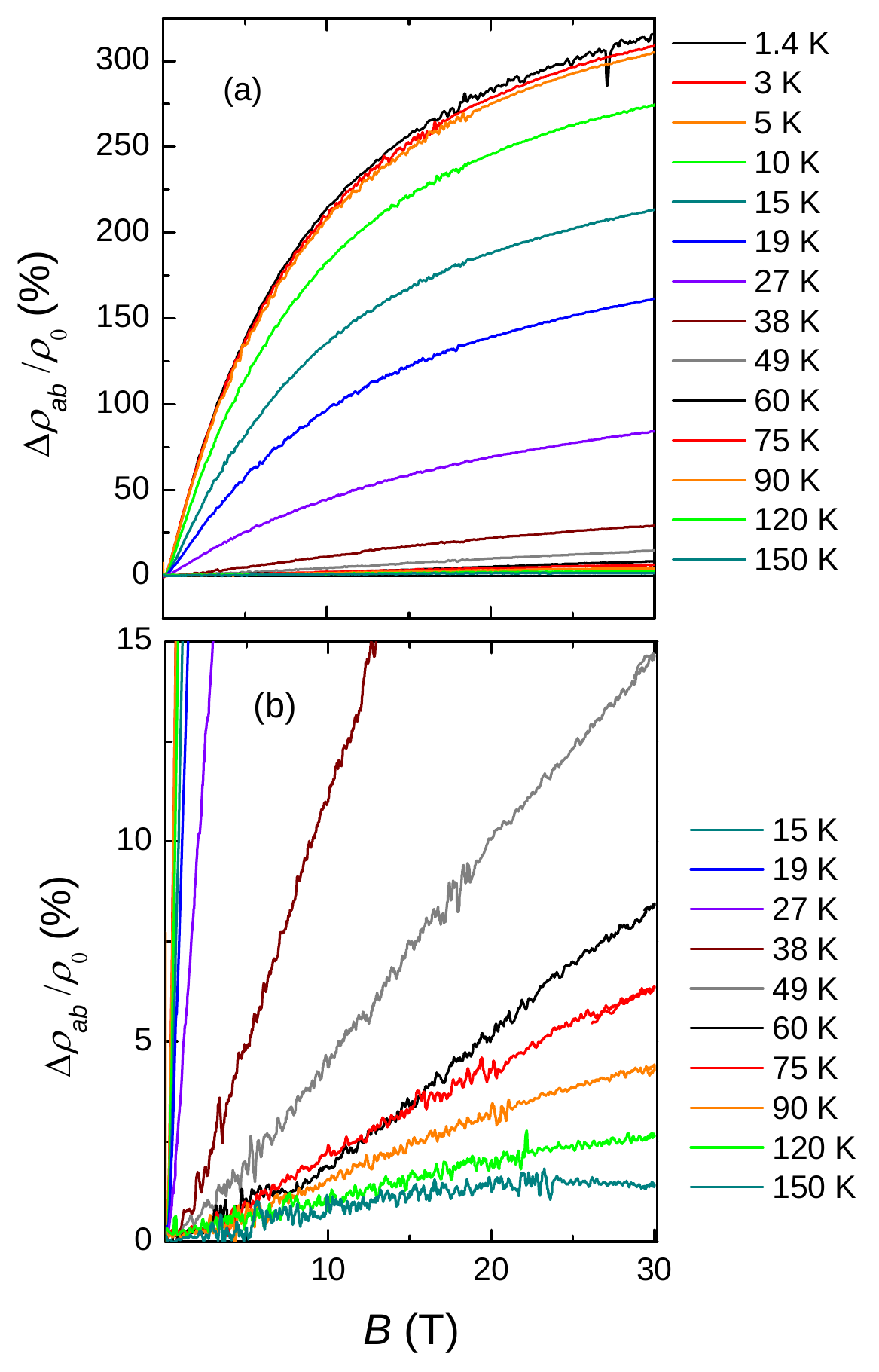}
\caption{(Color online). (a) Magnetic field dependence of the in-plane resistivity $\Delta \rho_{ab}(B)/\rho_0$ $(\rho_0$ = $\rho_{ab}(B=0))$ of PdCrO$_2$ in the temperature range from 1.4 K to 150 K in the field up to $B$ = 30 T perpendicular to the $ab$-plane. (b) The magnetoresistivity data shown in (a) on an expanded scale in order to emphasize the high-$T$ behavior.}
\label{fig:S1}
\end{figure}

\begin{figure}
\centering
\includegraphics[width=1.0\linewidth]{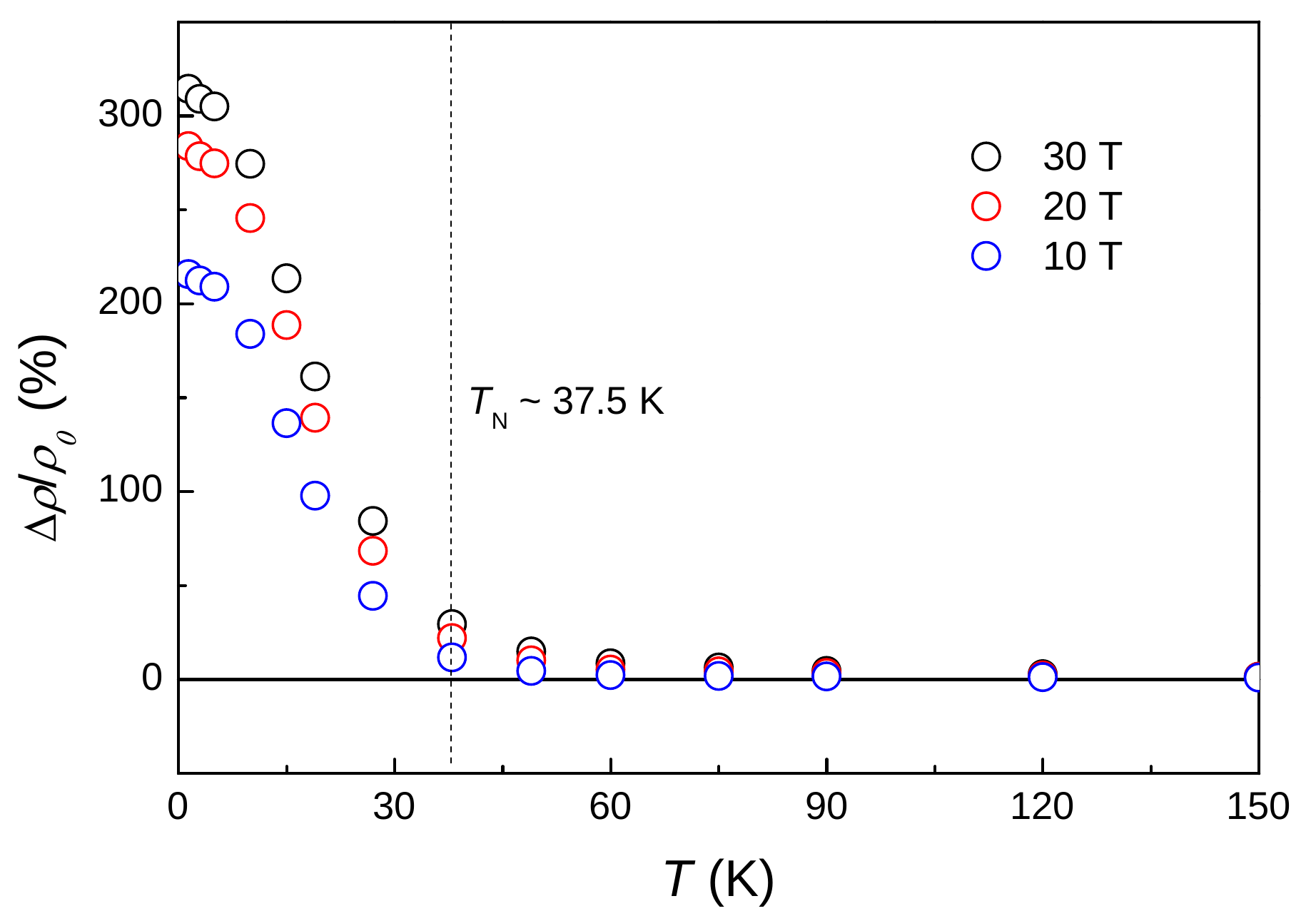}
\caption{(Color online). Temperature dependence of the magnetoresistivity of PdCrO$_2$ presented in Fig.~\ref{fig:S1} for constant values of magnetic field of 10 T, 20 T and 30 T. The antiferromagnetic ordering at $T_N$ is indicated by the dashed line.}
\label{fig:S2}
\end{figure}

\textbf{High-field magnetoresistivity in PdCrO$_2$.} Fig.~\ref{fig:S1} shows the magnetoresistivity (MR) $\rho_{ab}(B)$ =$\rho_0$ ($\rho_0 $ = $\rho_{ab}(B$ = 0)), recorded on a second crystal at fixed temperatures between 1.4 K and 150 K. In contrast to the magnetothermopower (MTP) behavior, the size of the MR decreases sharply with increasing temperature (see Fig.~\ref{fig:S2}). In the temperature range 95 K $\leq T \leq$ 130~K where the suppression of the thermopower reaches 100~\% in a field of 30 T, the MR is less than 5~\%, compared with 300~\% at $T$ = 1.4~K. 

The presented magnetoresistivity data was obtained as a part of the study published in Ref.~\cite{Ok}. There, the large transverse MR was reported together with a large longitudinal MR, indicating a significant coupling between itinerant electrons with the Cr spins.

\textbf{Determination of the effective mass of the $\alpha$ pocket.} Below $T_N$, where the Cr$^{3+}$ spins order, $S_{ab}/T$ increases as the conduction electrons undergo FS reconstruction~\cite{Ok, Hicks}. The existence of small pockets is confirmed by the observation of QO in the high-field thermopower below $T_N$. The observation of QO here illustrates not only the quality of the single crystals used in this study, but also the high sensitivity and stability of our optimized set-up.

\begin{figure}
\centering
\includegraphics[width=1.0\linewidth]{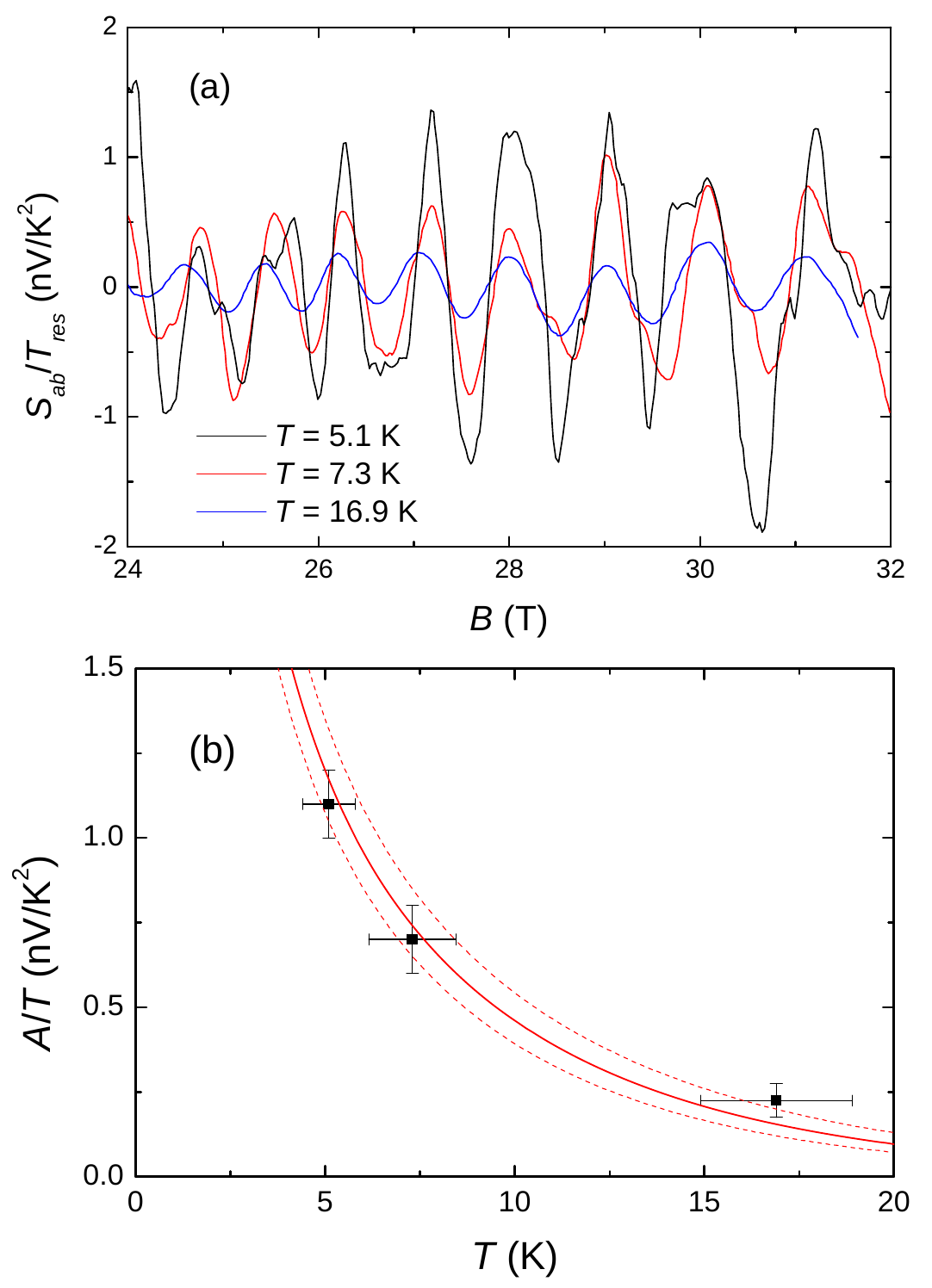}
\caption{(Color online). 
(a) The QO seen in $S_{ab}/T$ show suppression in amplitude with increasing temperature which satisfies the Lifshitz-Kosevich formalism \cite{Shoenberg}. (b) Temperature dependence of the thermopower oscillation amplitude $A/T$. The L-K fit is shown while the fit uncertainty is indicated with dotted lines. The $T$-error bars originate from a large applied thermal gradient, which is necessary for the detection of the quantum oscillations.}
\label{fig:S3}
\end{figure}

The $T$-dependence of the QO amplitude in $S_{ab}/T$ is presented in Fig.~\ref{fig:S3}a. Figure~\ref{fig:S3}b shows the Lifshitz-Kosevich (L-K) fit to the thermopower quantum oscillation amplitude determined over the field range 24-32 T. The fit to the L-K formula $A/T = [\sinh (am^{*}T/B)]^{-1}$ results in an effective mass of $m^* = 0.31(3) m_0$, where $m_0$ is the free electron mass.

\textbf{$S(T)$ vs. magnetic field}

\begin{figure}
\centering
\includegraphics[width=1.0\linewidth]{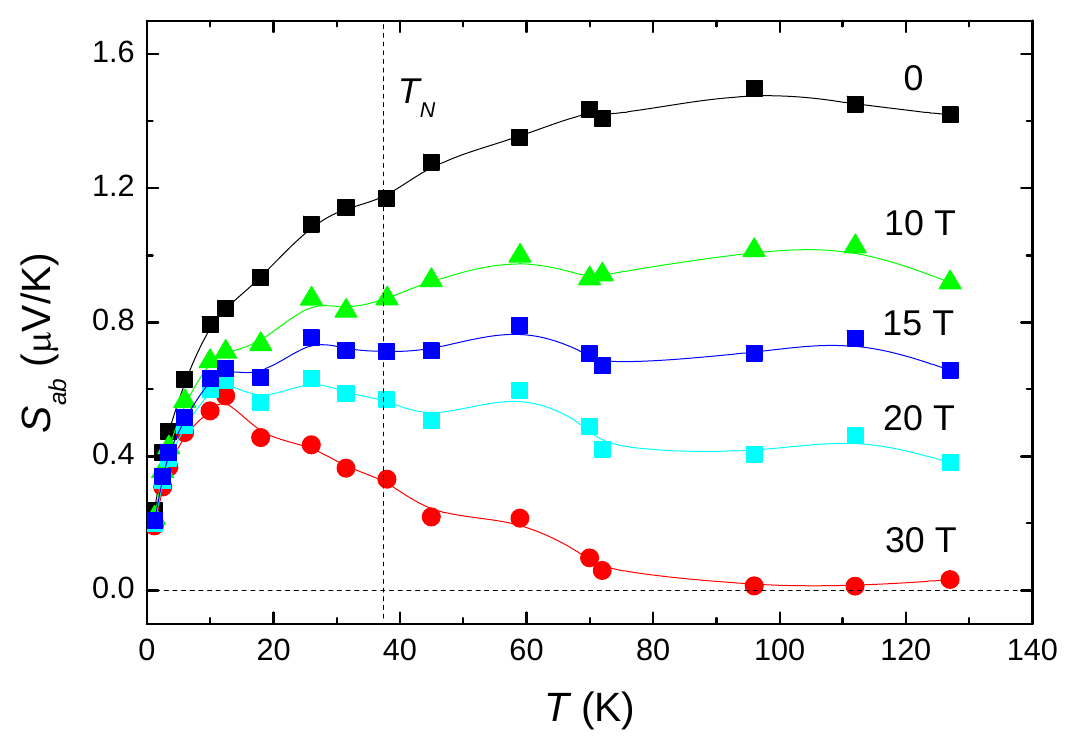}
\caption{(Color online). $S_{ab}$ as a function of temperature for values of magnetic field ranging from zero to $B$ = 30 T.}
\label{fig:S4}
\end{figure}

In the main text, Fig. 1 emphasizes the drop in the magnitude of the field-suppression of thermopower as the temperature is lowered and the magnetic phase is approached. The peak in $S_{ab}$ around 15~K for $B$ = 30 T, at first glance unconventional, is in fact an artefact of the plotting and the scatter in the data. This is most clearly seen in the plot of $S/T$ (Fig. 3 of original manuscript) where any putative feature at 15~K is lost. Fig.~\ref{fig:S4} shows the evolution of $S(T)$ for different field strengths. Here, it can be seen clearly that the peak at 15~K is indeed a by-product of the scatter in the data and, at $B$ = 30 T, is an consequence of the marked enhancement in $S/T$ with decreasing temperature on the one hand, and the thermodynamic requirement for the thermopower to vanish in the zero temperature limit on the other.

\textbf{Thermopower vs. neutron scattering.}

\begin{figure}
\centering
\includegraphics[width=1.0\linewidth]{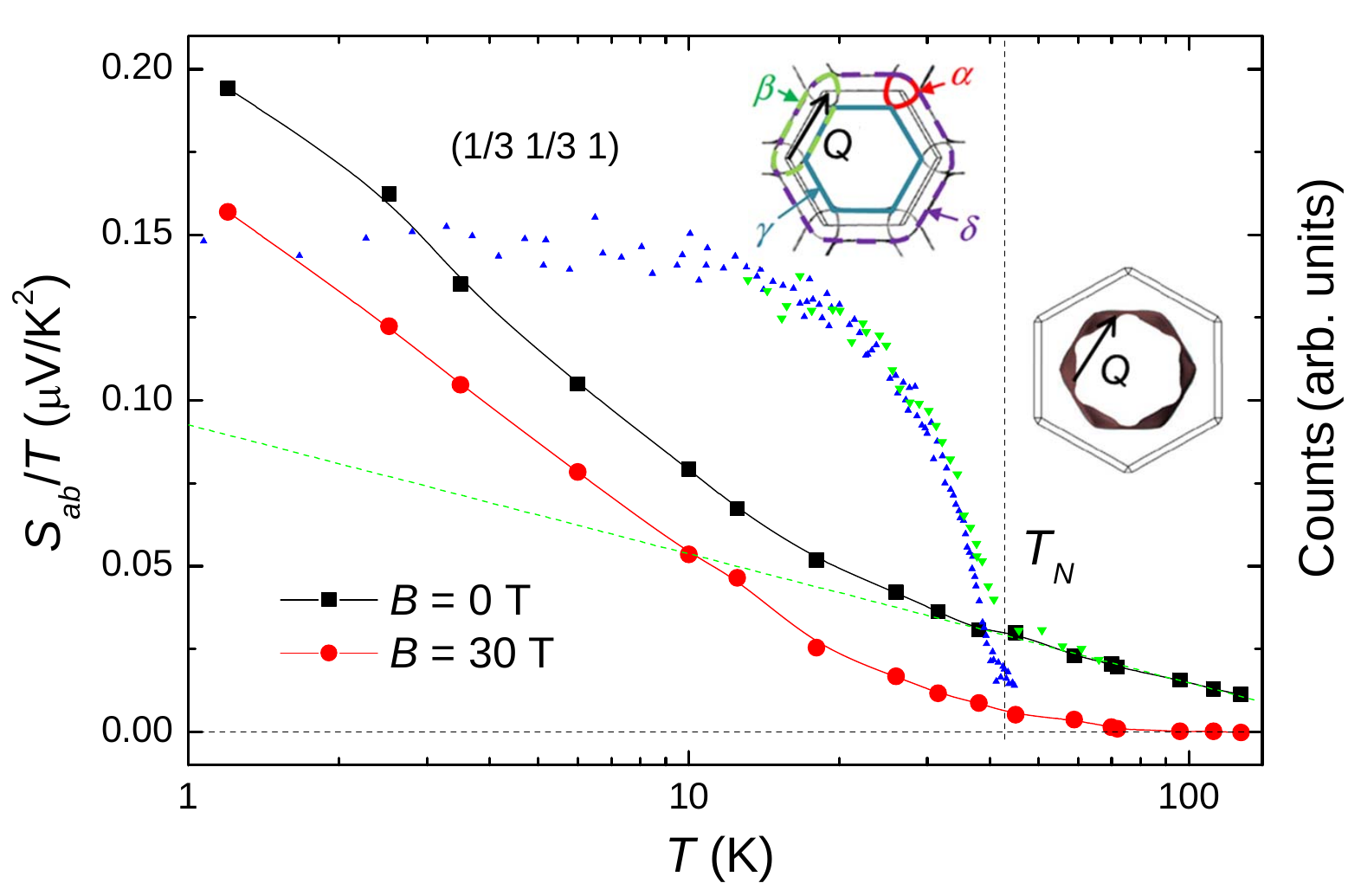}
\caption{(Color online). Temperature dependence of the neutron diffraction peak intensity at the $q=(\frac{1}{3},\frac{1}{3},1)$ position, as reported in Ref.~\cite{Mekata,TakatsuPRB}. The data is compared with the thermopower $S_{ab}/T$ data on a semilogarithmic scale presented in Fig.~3 of the main text.}
\label{fig:SN}
\end{figure}

The role of spin fluctuations in PdCrO$_2$ is underlined in Fig.~\ref{fig:SN} with a comparison of neutron scattering data from Ref.~\cite{Mekata,TakatsuPRB} with the thermopower $S_{ab}/T$ data shown in Fig.~3 of the main text. The scattering of itinerant electrons on short-range magnetic order above $T_N$ results in a linear $S/T$ behavior on a logarithmic temperature scale. Below $T_N$, formation of the long-range magnetic order results in a stronger peak intensity of neutron scattering which increases with lowering temperature. Coincidentally, the thermopower coefficient $S/T$ changes its slope at $T_N$, confirming a strong interaction of itinerant electrons with magnetic order.

\textbf{The possible role of magnon-electron interactions in the thermopower of PdCrO$_2$.} For the case of purely elastic scattering of electrons the thermoelectric power $S$ is expressed by Mott-Jones formula (see, e.g. Ref.~\cite{Mott}):

\begin{equation}\label{eq:Ssigma}
S\sigma = \frac{1}{e}\int dE \sigma(E) \frac{E-\mu}{T} \frac {\partial f}{\partial E}
\end{equation}

where $\mu$ is chemical potential, $f(E)$ is the Fermi function, $\sigma (E)$ is the conductivity at zero temperature and $\mu = E$ and $\sigma = -\int dE \sigma (E) \frac{\partial f}{\partial E}$ is a conductivity at a given temperature. If a typical energy scale of the dependence of $\sigma(E)$, $E_0$, is much larger than the temperature $T$, thermoelectric power is estimated as $S\approx \frac{k_B^{2}T}{eE_0}$, that is small (in a factor $\frac {k_B T}{E_0}$ smaller than a classical value $\frac{k_B}{e}$) and linear in temperature. Exceptions can be in the cases where electron density of states has unusually sharp energy dependence near the Fermi energy $E_F = \mu(T=0)$ or for Kondo systems~\cite{Kondo, Irkhin}. The former option is not applicable to PdCrO$_2$ according to the electronic structure calculations~\cite{Ong}. The latter case would assume the formation of the Kondo lattice state. Although this can not be completely excluded it seems less probable for a 4$d$ system. Also, the experimental data on heat capacity and magnetic susceptibility do not support this scenario~\cite{Takatsu}. 
%This option should be probably kept for future more detailed investigations but currently we would prefer to find a more conventional explanation. 

It is known that when inelastic scattering mechanisms are relevant such as electron-phonon scattering, the nonequilibrium effects in the scatterer subsystem such as phonon drag can be of crucial importance for the thermoelectric power~\cite{Gurevich, Baylin, Ziman}. It is natural to assume that in magnetic systems magnons (spin waves) can play a similar role (for the case of ferromagnets, see e.g. References~\cite{Bhandari, Grannemann, Costache}. Keeping in mind that the measured $S$ (see Fig.~1) is essentially nonlinear in temperature and very strongly dependent on magnetic field (except the case of low temperatures) it seems to be natural to attribute these features to magnon drag. The latter is relevant if the magnon-electron scattering is, at least, comparable (or more important) than the scattering on magnons by defects. This assumption looks very reasonable for PdCrO$_2$ which is a system with a low level of defects.

PdCrO$_2$ is a quasi-two-dimensional itinerant-electron antiferromagnet with the Neel temperature $T_N =$ 37.5~K much smaller than a typical energy of exchange interactions $J$, of the order of paramagnetic Curie-Weiss temperature $\Theta_W =$ 500 K~\cite{Takatsu}. Importantly, a short-range magnetic order remains strong enough, with a correlation length $\xi \gg a$ ($a$ is the lattice period) up to the temperatures $T \approx J \gg T_N$ (see Fig.~\ref{fig:SN}). As was shown in Ref.~\cite{Irkhin91}, character of electron-magnon interactions in such a case does not change essentially at $T = T_N$ and remains basically the same assuming that $\xi > k_F^{-1}$ which in the case of metals corresponds to $T < J$. Therefore, we can use until these temperatures a theory of electron-magnon interaction in itinerant-electron antiferromagnets developed in Ref.~\cite{Irkhin00}.

In isotropic antiferromagnets, in the absence of magnetic field, the magnon frequency $\omega_{\vec{q}}$ tends to zero linearly for the wave vector $\vec{q} \rightarrow 0$ and $\vec{q} \rightarrow \vec{Q}$ where $\vec{Q}$ is the antiferromagnetic order wave vector; for the theory specifically describing magnons in a triangular lattice with the nearest-neighbor antiferromagnetic interactions please refer to Ref.~\cite{Chernyshev}. There is a dramatic difference in the character of electron-magnon interactions with these two types of soft magnons~\cite{Irkhin00}: for $\vec{q} \rightarrow 0$, the scattering probability with the momentum-transfer $\vec{q}$, $W_{vec{q}} \propto \omega_{\vec{q}} \propto q$ vanishes (identically to the scattering of electrons by acoustic phonons~\cite{Ziman}) whereas for $\vec{q} \rightarrow \vec{Q}$ it is formally divergent: $W_{\vec{q}} \propto 1/\omega_{\vec{q}} \propto |\vec{q}-\vec{Q}|^{-1}$. These arguments indicate that the latter class of scattering processes is much more relevant. Simultaneously, this process requires scattering to overcome the antiferromagnetic energy gap $\Delta$ and therefore there is a threshold in temperature: such strong electron-magnon interaction becomes relevant at $T > T^{*} \approx (\Delta/E_F)J$. We believe that these arguments provide a qualitative understanding of the MTP behavior shown in Fig.~1.

The Hamiltonian of antiferromagnet in the spin-wave temperature region ($T \ll J$) is formally equivalent to that for electron-phonon interaction (with different matrix elements). To estimate the drag contribution $S_g$ to the thermoelectric power one can use the expression for the phonon case~\cite{Baylin}:

\begin{equation}\label{eq:Sg}
S_g=\frac{1}{|e|\left\langle (\nu^e_{\vec{k},\chi})^2\right\rangle}\sum_{\vec{q}} \frac{\partial N_{\vec{q}}}{\partial T}\alpha_{\vec{q}}\nu^m_{\vec{q}\chi} \left\langle \nu^{\epsilon}_{\vec{k}+\vec{q},\chi}-\nu^{\epsilon}_{\vec{k},\chi} \right\rangle
\end{equation}

where $N_{\vec{q}}=\left[\exp(\hbar\omega_{\vec{q}}/k_BT)-1) \right]^{-1}$ is Bose distribution function, $\nu^m_{\vec{q} \chi}=\partial \omega_{\vec{q}}/\partial q_{\chi}$ is magnon group velocity along current propagation $x$-direction,  $\nu^{\epsilon}_{\vec{k}, \chi}$ is electron group velocity in the same direction, angular brackets means the average over the Fermi surface for the vector $\vec{k}$, $\alpha_{\vec{q}}$ is the rate of scattering of magnon with the wave vector $\vec{q}$ by electrons normalized to the total probability of scattering of this phonon due to all interaction where this magnon is involved (namely, with defects, sample boundaries, and magnon-magnon interactions). For example, if magnons are in almost ballistic regime, one can estimate $\alpha_{\vec{q}}\approx W_{\vec{q}}L/\nu^m_{\vec{q}}$, where $L$ is the sample size. 

Below $T^*$, where “singular” electron-magnon scattering processes with $\vec{q} \rightarrow \vec{Q}$ are forbidden, the contribution from small $q$ is dominant. One has $\alpha_{\vec{q}}\propto q$, $\nu^m_{\vec{q}\chi}\left\langle \nu^{\epsilon}_{\vec{k}+\vec{q},\chi}-\nu^{\epsilon}_{\vec{k},\chi}\right\rangle \propto q$, and $S_g \propto \frac{\partial}{\partial T}\sum_{\vec q}N_{\vec{q}}q^2 \propto T^3$. In this regime, the Mott-Jones contribution~\cite{Mott} is dominant, thermoelectric power is roughly proportional to temperature and weakly dependent on magnetic field. 

Above $T^*$, where “singular” electron-magnon scattering processes with $\vec{q} \rightarrow \vec{Q}$ are allowed, the contribution from small $q$ is dominant. One has $\alpha_{\vec{q}}\propto|\vec{q}-\vec{Q}|^{-1}$, $\nu^m_{\vec{Q},\chi}\left\langle \nu^{\epsilon}_{\vec{k}+\vec{Q},\chi}-\nu^{\epsilon}_{\vec{k},\chi}\right\rangle$ is constant, and $S_g\propto \frac{1}{T^2}\sum_{|\vec{q}-\vec{Q}|>q^*} \frac{\omega_{\vec{q}}N_{\vec{q}}(1+N_{\vec{q}})}{\omega_{\vec{q}}} \propto \ln T/T^*$, where $q^* \approx \Delta/\hbar\nu_{F}$ is the infrared cut-off wave vector~\cite{Irkhin00}. In this regime, the drag contribution is dominant. Indeed, temperature dependence of thermoelectric power in the absence of magnetic field at relatively high temperatures is roughly logarithmic. One can estimate $T^*$ as 20-50 K which seems to be a reasonable estimate for $J \approx$ 150-500 K, $\Delta/E_F\leq$~0.1. Indeed, from the electronic structure calculations for ferromagnetic phase~\cite{Ong} the splitting between spin-up and spin-down Pd states $\Delta \leq$ 1 eV, $E_F \approx$ 8 eV.
The drag contribution to the $S$ is strongly dependent on the spin-wave spectrum which is very sensitive to magnetic field~\cite{Gekht, Starykh}. This may explain qualitatively the dramatic growth of magnetothermoelectric power at higher temperatures.